\documentclass[pdflatex,sn-mathphys-num]{sn-jnl}

\usepackage[utf8]{inputenc}
\usepackage{graphicx}%
\usepackage{multirow}%
\usepackage{amsmath,amssymb,amsfonts}%
\usepackage{amsthm}%
\usepackage{mathrsfs}%
\usepackage[title]{appendix}%
\usepackage{xcolor}%
\usepackage{textcomp}%
\usepackage{manyfoot}%
\usepackage{booktabs}%
\usepackage{algorithm}%

\usepackage{algorithmicx}%
\usepackage{algpseudocode}%
\usepackage{listings}%
\usepackage{adjustbox}

\theoremstyle{thmstyleone}%
\theoremstyle{thmstyletwo}%

\theoremstyle{thmstylethree}%

\begin{document}

\title[Article Title]{MatMind: A Structure-Activity Knowledge-Driven Generative Foundation Model for Materials Science}


\author[1,2]{\fnm{Zhan'ao} \sur{Yao}}
\equalcont{These authors contributed equally to this work}
\author[4,5]{\fnm{Boxuan} \sur{Zhang}}
\equalcont{These authors contributed equally to this work}

\author[6,7]{\fnm{Jingyuan} \sur{Shu}}
\equalcont{These authors contributed equally to this work}
\author[6]{\fnm{Xiaoyu} \sur{Wu}}  
\equalcont{These authors contributed equally to this work}
\author[1,2]{\fnm{Rongyan} \sur{Wang}}

\author[4,5]{\fnm{Linjing} \sur{Li}}

\author[4,5]{\fnm{Dajun} \sur{Zeng}}

\author[7]{\fnm{Yudong} \sur{Yao}}
\author*[8]{\fnm{Tingwei} \sur{Chen}}\email{twchen@lnu.edu.cn}
\author*[1,2]{\fnm{Youwei} \sur{Wang}}\email{ywwang@mail.sic.ac.cn}

\author*[1,2]{\fnm{Xiaolin} \sur{Zhao}}\email{zhaoxiaolin@mail.sic.ac.cn}
\author*[4,5]{\fnm{Jiahui} \sur{Shi}}\email{jiahui.shi@ia.ac.cn}

\author*[1,2,3]{\fnm{Jianjun} \sur{Liu}}\email{jliu@mail.sic.ac.cn}

\affil[1]{\orgdiv{State Key Laboratory of High Performance Ceramics}, \orgname{Shanghai Institute of Ceramics, Chinese Academy of Sciences}, \orgaddress{\street{1295 Dingxi Road}, \city{Shanghai}, \postcode{200050}, \country{China}}}

\affil[2]{\orgdiv{Center of Materials Science and Optoelectronics Engineering}, \orgname{University of Chinese Academy of Sciences}, \orgaddress{\city{Beijing}, \postcode{100049}, \country{China}}}

\affil[3]{\orgdiv{School of Chemistry and Materials Science, Hangzhou Institute for Advanced Study}, \orgname{University of Chinese Academy of Sciences}, \orgaddress{\street{1 Sub-lane Xiangshan}, \city{Hangzhou}, \postcode{310024}, \country{China}}}
\affil[4]{\orgdiv{State Key Laboratory of Multimodal Artificial Intelligence Systems}, \orgname{Institute of Automation, Chinese Academy of Sciences}, \orgaddress{\street{95 Zhongguancun East Road}, \city{Beijing}, \postcode{100190}, \country{China}}}

\affil[5]{\orgdiv{School of Artificial Intelligence}, \orgname{University of Chinese Academy of Sciences}, \orgaddress{\city{Beijing}, \postcode{100049}, \country{China}}}

\affil[6]{\orgname{Beijing Wenge Technology Co., Ltd.} , \street{Room 717, 7th Floor, Building 9, No. 9 West Beisihuan Road}, \city{Beijing}, \postcode{100080}, \state{Beijing},\country{China}}

\affil[7]{\orgdiv{College of Medicine and Biological Information Engineering}, \orgname{Northeastern University}, \orgaddress{\street{No. 3-11, Wenhua Road, Heping District}, \city{Shenyang}, \postcode{110819}, \state{Liaoning}, \country{China}}}

\affil[8]{\orgdiv{Faculty of Information}, \orgname{Liaoning University}, \orgaddress{\street{66 Chongshan Middle Road}, \city{Huanggu Qu}, \postcode{110031}, \state{Shenyang}, \country{China}}}

\abstract{Progress in AI-driven crystal materials science has so far been carried by narrow architectures purpose-built for individual tasks---graph neural networks for property prediction, diffusion and flow-matching models for crystal generation---each excelling within its niche yet unable to act as a shared backbone across the full spectrum of materials problems. Generative large language models offer a fundamentally different paradigm, in which structural representation, quantitative prediction, and structure--activity reasoning can be unified within one model, but the materials community has yet to see this paradigm realized at a level competitive with established narrow specialists. Here we present MatMind, a generative foundation model purpose-built for crystal materials science under this paradigm, developed through the coordinated activation of structure--activity knowledge and physics-informed feedback within a progressive training framework---combining structure--activity knowledge injection, a dual-head architecture that jointly trains language reasoning and numerical regression in a shared representation space, and multi-objective physics-informed reinforcement learning over stability, novelty, and structural diversity. Across three task families, MatMind attains the lowest mean absolute error on energy above hull, bulk modulus, and band gap---surpassing graph neural network predictors purpose-built for these tasks---reaches an S.U.N.\ rate of 65.3\% on unconditional crystal generation, and achieves a comparable multiplicative improvement on magnetization-density-conditioned generation, where only 21 positive samples exist within over 600{,}000 training entries. By matching or surpassing narrow specialists on their own ground while operating within a single unified model, MatMind shows that the LLM-based paradigm can serve as a viable backbone for crystal materials science going forward.}




\maketitle

\section{Introduction}\label{sec1}

The use of generative large language models in materials science has rapidly grown into a distinct research direction~\cite{TODO_llm_for_materials_review,TODO_materials_foundation_models}, motivated by the natural fit between such models and the central representational needs of the field: crystal structures admit a natural sequential encoding~\cite{TODO_crystal_sequence_representation}, the underlying physics provides well-defined quantitative targets~\cite{TODO_materials_property_prediction}, and a large body of structure--property knowledge is already available in textual form~\cite{TODO_materials_text_mining,TODO_scientific_language_models}. Building on this fit, a series of materials-oriented language models has been developed, coupling generative large language models to crystal generation, property prediction, and related tasks, and providing initial evidence that the paradigm is viable~\cite{TODO_llm_prop,TODO_materials_llm_generation,TODO_matbert}. Yet the capability actually realized so far falls well short of what such materials-oriented language models could in principle support. Thermodynamic plausibility is rarely enforced as a structural property of the model~\cite{TODO_crystal_generation_limitations}, quantitative prediction is typically treated as a problem separate from language reasoning rather than as a coupled one~\cite{TODO_llm_prop}, and the underlying structure--activity understanding on which both tasks depend has not been systematically built into the model~\cite{TODO_structure_activity_materials}. The result is a class of materials-oriented language models that succeed task by task but do not yet behave as genuinely materials-specialized foundation models. Closing this gap is the central challenge the present work addresses.

Prior work has approached this challenge from several directions, yet each trajectory faces fundamental limitations intrinsic to its design. Autoregressive generation methods bring crystal structures into the language modeling framework but lack genuine physical-world constraints, leaving the thermodynamic plausibility of generated structures unguaranteed~\cite{TODO_materials_llm_generation,TODO_physical_validity_crystal_generation}. Physics-informed reinforcement learning approaches attempt to ground generation in real physical feedback~\cite{TODO_physics_informed_rl,TODO_rl_for_materials_generation}, but the foundation models serving as optimization starting points lack sufficient materials science prior, and reward designs remain immature in achieving genuine coordination across stability, novelty, and structural diversity~\cite{TODO_multiobjective_materials_generation}. For property prediction, existing approaches treat language understanding and numerical prediction as separate problems~\cite{TODO_llm_prop,TODO_matbert,TODO_gnn_crystal_property_prediction}, overlooking their potential synergy and limiting the depth of structure--activity understanding the model can develop. These shared difficulties indicate that bridging the gap between general linguistic knowledge and specialized materials science capability requires a fundamental rethinking of the training paradigm rather than incremental refinement along any existing direction.

Here we present MatMind, a materials-specialized generative foundation model whose core design principle is the coordinated activation of structure--activity knowledge and physics-informed reinforcement learning. MatMind addresses each of these limitations through a progressive training framework: crystal scientific data alignment pretraining and structure--activity relationship enhanced fine-tuning establish the materials science prior that existing approaches lack; a dual-head architecture jointly trains language reasoning and numerical regression within a shared representation space, allowing the two to mutually reinforce rather than be optimized in isolation; and multi-objective physics-informed reinforcement learning guides physically plausible crystal generation under coordinated optimization of stability, novelty, and structural diversity. Across these components, structure--activity knowledge and physical feedback are designed to operate in a mutually reinforcing manner---the former providing scientific prior for the latter, and the latter anchoring the former in verifiable physical reality.

We evaluate MatMind across three task families that together cover the central capabilities of a materials-specialized generative foundation model. On quantitative property prediction, MatMind achieves the lowest mean absolute error among all evaluated methods on the three benchmark tasks of energy above hull, bulk modulus, and band gap, outperforming both graph neural network models purpose-built for crystal property prediction (CGCNN, M3GNet)~\cite{TODO_cgcnn,TODO_m3gnet} and language-model-based predictors (LLM-Prop, MatBERT-109M)~\cite{TODO_llm_prop,TODO_matbert}. On unconditional crystal generation, MatMind attains an S.U.N.\ rate of 65.3\%, taking a leading position among the compared diffusion- and language-model-based baselines (MatterGen, DiffCSP)~\cite{TODO_mattergen,TODO_diffcsp}. On conditional generation and small-data property transfer, physics-informed reinforcement learning further drives the generated distribution toward target property intervals across band gap, bulk modulus, and---most strikingly---magnetization density, where only 21 positive samples are available within over 600{,}000 training entries, yet a comparable multiplicative improvement in target-property generation is still achieved. This last result demonstrates that the framework remains effective in a regime where conventional supervised approaches are fundamentally underdetermined~\cite{TODO_low_data_materials_discovery}, opening a scalable and general path for accelerating computational discovery of functional materials whose target properties are represented by only a small number of known examples.

\section{Result}\label{sec2}
\begin{figure}[htp]
    \centering
    \begin{adjustbox}{center}
        \includegraphics[width=20cm]{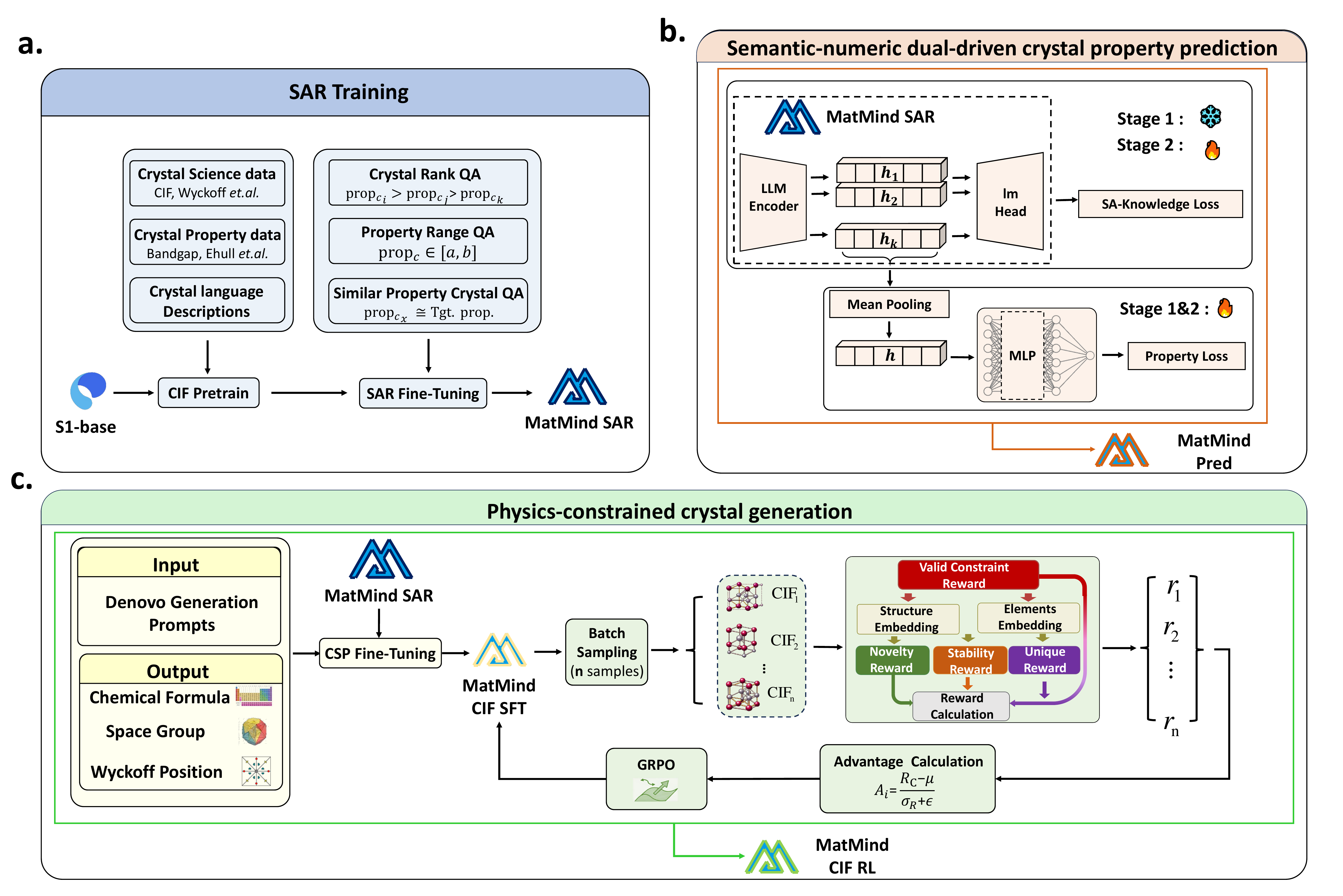}
    \end{adjustbox}
    \caption{Overview of the MatMind framework. \textbf{a}, Stage~1: foundation model
construction. Crystal scientific data alignment pretraining on randomly interleaved CIF,
property-annotation, and natural-language descriptions, followed by structure--activity
relationship (SAR) enhanced fine-tuning over crystal ranking, property-interval prediction,
and target-guided selection tasks with chain-of-thought supervision, yielding MatMind~SAR.
\textbf{b}, Stage~2: semantic--numeric dual-driven property prediction. A dual-head
architecture jointly trains a language head (SA-knowledge loss) and a numerical regression
head (property loss) within a shared representation space via mean pooling of final hidden
states. \textbf{c}, Stage~3: physics-constrained crystal generation. MatMind~SAR is supervised
fine-tuned on de novo Wyckoff-sequence generation and then optimized with GRPO under a multi-objective physics-informed reward combining validity, stability, novelty, and diversity.}
    \label{fig:Method}
\end{figure}

\subsection{MatMind framework}

MatMind builds upon S1-Base 8B as the base model~\cite{TODO_s1_base} and systematically develops specialized materials science capability through a three-stage progressive training pipeline (Fig.~1a--1c). The first stage is foundation model construction (Fig.~1a). MatMind begins with crystal scientific data alignment pretraining on a large-scale crystal-text corpus sourced from the thermodynamically stable subset of the Alexandria database~\cite{TODO_alexandria_database}, with three data categories---crystallographic information file (CIF) representation~\cite{TODO_cif_format}, physical property annotations, and natural language crystal descriptions---arranged in randomly interleaved sequences. This design enables the model to naturally establish internal associations among crystal structures, physical properties, and textual descriptions while learning each type of crystallographic knowledge, rather than memorizing each category in isolation. Pretraining establishes foundational scientific understanding of key crystallographic concepts---including space group symmetry, Wyckoff positions, and thermodynamic stability~\cite{TODO_space_group_symmetry,TODO_wyckoff_positions,TODO_thermodynamic_stability}---bridging the knowledge gap caused by the extreme sparsity of crystallographic signal in general-purpose corpora~\cite{TODO_scientific_corpus_gap}. Structure-activity relationship enhanced fine-tuning across three task types---crystal performance ranking, performance interval prediction, and target-guided crystal selection---then uses chain-of-thought reasoning as an intermediate bridge connecting task instructions to target answers~\cite{TODO_chain_of_thought}, elevating the model's understanding of structure-activity relationships from implicit textual induction to explicit causal reasoning and yielding the MatMind SAR model.

The second stage is predictive model construction (Fig.~1b). Building on MatMind SAR, a dual-head architecture jointly training a language head and a numerical regression head is introduced through a two-step strategy: the first step freezes the LLM backbone for regression head warm-up, and the second step unfreezes all parameters for joint training under a unified loss function that simultaneously optimizes language reasoning and quantitative property prediction~\cite{TODO_multitask_learning,TODO_joint_language_regression}. The numerical regression head performs direct continuous-value prediction of properties such as band gap and bulk modulus through mean pooling and linear transformation of final hidden states, bypassing the inherent precision limitations of tokenization~\cite{TODO_joint_language_regression,TODO_tokenization_numerical_precision}; the language head is supervised by structure-activity reasoning distillation data~\cite{TODO_reasoning_distillation}, enabling the model to explicitly output causal understanding of structure-property relationships in natural language form. The coordinated training of both heads within a shared representation space enables mutual reinforcement between language reasoning and quantitative prediction capabilities, elevating structure-activity understanding from statistical fitting to causal reasoning.

The third stage is generative model construction (Fig.~1c). MatMind SAR undergoes supervised fine-tuning on de novo instruction samples using Wyckoff representation as a compact text encoding for crystal structures~\cite{TODO_wyckoff_positions,TODO_crystal_sequence_representation}, preserving complete crystallographic information while substantially reducing sequence length and establishing the basic capability for crystal structure sequence generation from scratch. Physics-informed reinforcement learning via Group Relative Policy Optimization is then applied~\cite{TODO_grpo,TODO_physics_informed_rl} with a hierarchical four-objective reward framework spanning structural validity, thermodynamic stability, structural novelty, and compositional diversity guiding policy updates~\cite{TODO_multiobjective_materials_generation}. Structural validity serves as a hard gate restricting subsequent reward computation to structures passing interatomic distance, charge neutrality, and relaxation convergence checks~\cite{TODO_multiobjective_materials_generation,TODO_structure_relaxation}; stability, novelty, and diversity rewards are all mapped to a unified linear range and combined through summation into the final reward signal, a design that endows the reward framework with strong extensibility. The fundamental distinction from existing physics-informed reinforcement learning approaches lies in the starting point: MatMind's reinforcement learning begins from the materials science prior accumulated in the first stage, enabling physical reward signals to effectively guide policy optimization on a foundation of genuine scientific understanding---maintaining structural diversity while improving stability~\cite{TODO_rl_for_materials_generation,TODO_diversity_stability_tradeoff}.

\subsection{Semantic-numeric dual-driven crystal property prediction}
\begin{figure}[htp]
    \centering
    \begin{adjustbox}{center}
        \includegraphics[width=20cm]{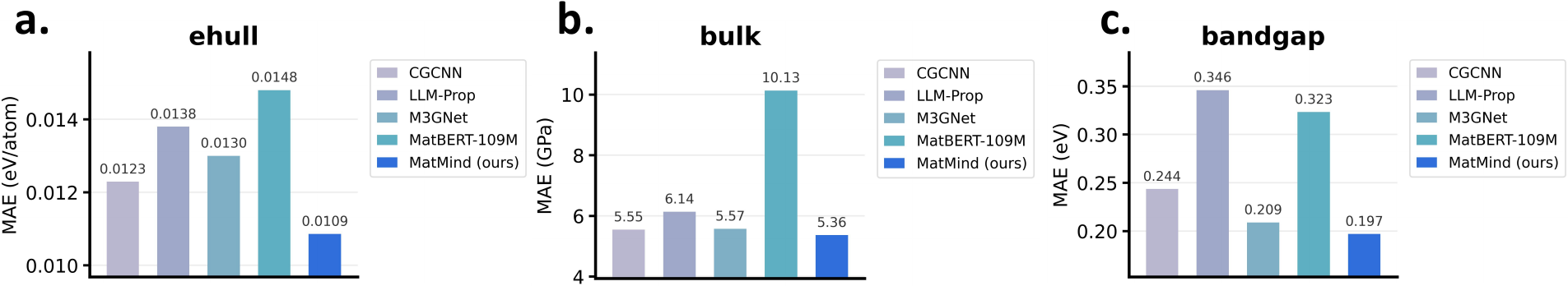}
    \end{adjustbox}
    \caption{Quantitative property prediction performance. Mean absolute error (MAE) of MatMind
against graph neural network predictors (CGCNN, M3GNet) and language-model-based predictors
(LLM-Prop, MatBERT-109M) on three tasks. \textbf{a}, Energy above the convex hull
($E_{\text{hull}}$, eV/atom). \textbf{b}, Bulk modulus (GPa). \textbf{c}, Band gap (eV). Lower
is better; MatMind attains the lowest or near-lowest MAE on all three tasks.}
    \label{fig:Method}
\end{figure}

MatMind's performance across three quantitative property prediction tasks---energy above the convex hull ($E_{\text{hull}}$), bulk modulus, and band gap---validates the effectiveness of structure-activity knowledge-driven training (Fig.~2). These three tasks span distinct physical dimensions of materials functionality: $E_{\text{hull}}$ reflects thermodynamic stability~\cite{TODO_thermodynamic_stability}, bulk modulus reflects mechanical response characteristics~\cite{TODO_bulk_modulus_materials}, and band gap reflects electronic structure properties~\cite{TODO_band_gap_electronic_structure}, together constituting a comprehensive test of MatMind's structure-activity understanding capability.

On the $E_{\text{hull}}$ prediction task (Fig.~2a), MatMind achieves a MAE of 0.0109~eV/atom, outperforming CGCNN (0.0123~eV/atom), LLM-Prop (0.0138~eV/atom), M3GNet (0.0130~eV/atom), and MatBERT-109M (0.0148~eV/atom)~\cite{TODO_cgcnn,TODO_llm_prop,TODO_m3gnet,TODO_matbert}. Notably, $E_{\text{hull}}$ prediction accuracy carries direct implications for the downstream crystal generation task---the stability reward signal in MatMind's reinforcement learning stage relies on MLIP energy evaluations~\cite{TODO_mlip_materials_stability}, and accurate $E_{\text{hull}}$ understanding enables more effective policy response to stability reward signals. On the bulk modulus prediction task (Fig.~2b), MatMind achieves a MAE of 5.36~GPa, comparable to CGCNN (5.55~GPa) and M3GNet (5.57~GPa) while substantially outperforming LLM-Prop (6.14~GPa) and MatBERT-109M (10.13~GPa). On the band gap prediction task (Fig.~2c), MatMind achieves a MAE of 0.197~eV, significantly surpassing CGCNN (0.244~eV), LLM-Prop (0.346~eV), M3GNet (0.209~eV), and MatBERT-109M (0.323~eV).

Across all three tasks, MatMind achieves the best or near-best prediction accuracy, with the comparison models spanning both graph neural network approaches (CGCNN, M3GNet)~\cite{TODO_cgcnn,TODO_m3gnet} and language model-based approaches (LLM-Prop, MatBERT-109M)~\cite{TODO_llm_prop,TODO_matbert}. This demonstrates that the structure-property representations established through MatMind's structure-activity knowledge injection generalize across property types rather than constituting task-specific local fitting. Particularly noteworthy is that MatMind, as a method built upon a 8B-parameter generative large language model, surpasses graph neural network models purpose-built for crystal property prediction in quantitative prediction accuracy, further validating the effectiveness of the three-stage progressive training strategy proposed in this work---the scientific reasoning potential of general-purpose large language models can be fully unlocked through systematic domain knowledge injection~\cite{TODO_domain_adaptation_llm_science,TODO_materials_foundation_models}.

\subsection{Physics-constrained crystal generation}
\begin{figure}[htp]
    \centering
    \begin{adjustbox}{center}
        \includegraphics[width=18cm]{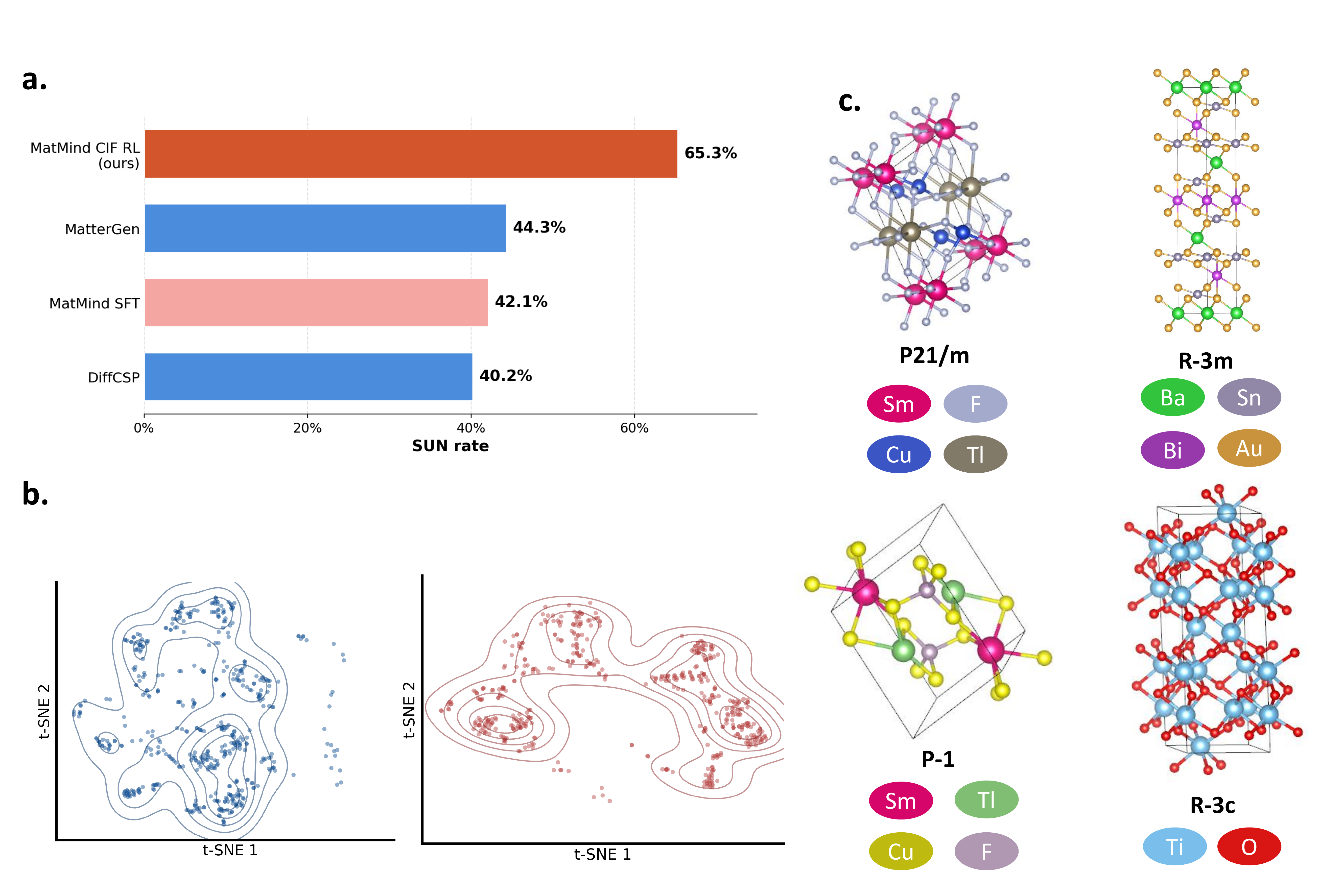}
    \end{adjustbox}
    \caption{Unconditional crystal generation. \textbf{a}, S.U.N.\ rate of MatMind~CIF~RL
(65.3\%) compared with MatterGen (44.3\%), DiffCSP (40.2\%), and the supervised-only ablation
MatMind~SFT (42.1\%), all evaluated under the identical relaxation, OMat energy-correction,
and MP-hull protocol. \textbf{b}, t-SNE visualization of generated structures in chemical
space for MatMind~SFT (left) versus MatMind~CIF~RL (right); reinforcement learning expands
exploration into regions far from the training distribution while retaining thermodynamic
stability. \textbf{c}, Representative generated structures spanning binary to quaternary
systems, including SmTlCu$_2$F$_8$, BaSn$_2$BiAu$_4$, LiSmPS$_4$, and a corundum-type Ti$_2$O$_3$
(R$\bar{3}$c) recovered despite the training set containing only metastable Ti$_2$O$_3$ polymorphs.}
    \label{fig:Method}
\end{figure}

\subsubsection{Unconditional crystal generation}
MatMind demonstrates substantial performance advantages on the unconditional crystal generation task (Fig.~3a). Using MatterGen and DiffCSP as comparison baselines~\cite{TODO_mattergen,TODO_diffcsp}, MatMind CIF RL achieves a S.U.N. rate of 65.3\%, representing an improvement of over 20 percentage points compared to MatterGen (44.3\%) and DiffCSP (40.2\%). This result demonstrates that, through systematic structure-activity knowledge injection and physics-informed reinforcement learning, large language models can serve as an effective paradigm for crystal structure generation~\cite{TODO_materials_llm_generation}, with the natural compatibility between their autoregressive generation framework and physical reward signals providing a unique advantage for real-time coupling of stability constraints with the generation process~\cite{TODO_rl_for_materials_generation,TODO_physics_informed_rl}.

A controlled experiment directly validates the necessity of physics-informed reinforcement learning (Fig.~3a). Replacing reinforcement learning with pure supervised fine-tuning (MatMind SFT) causes the S.U.N. rate to drop from 65.3\% to 42.1\%---a decrease of over 23 percentage points---demonstrating that supervised fine-tuning establishes basic generative capability but cannot spontaneously produce high proportions of thermodynamically stable novel structures without explicit physical constraint guidance~\cite{TODO_physical_validity_crystal_generation,TODO_thermodynamic_stability}. Notably, the S.U.N. rate of MatMind SFT (42.1\%) is comparable to that of MatterGen (44.3\%), implying that the performance gains contributed by the reinforcement learning stage essentially constitute the entirety of MatMind's advantage over diffusion-based methods, further highlighting the critical role of real-time physical reward signal guidance during the generation process.

t-SNE visualization of generated structure distributions in chemical space further reveals the deep impact of reinforcement learning on generative behavior (Fig.~3b)~\cite{TODO_tsne}. Comparing the generation distributions of MatMind SFT and MatMind CIF RL, the introduction of physics-informed reinforcement learning significantly expands the model's exploration into unknown chemical space regions, with generated structures extending substantially into areas far from the training distribution, while structures in these newly explored regions maintain high thermodynamic stability---indicating that reinforcement learning does not simply direct generation toward known stable configurations, but rather endows the model with the ability to actively discover novel stable structures in unfamiliar chemical space~\cite{TODO_chemical_space_exploration,TODO_diversity_stability_tradeoff}.

Fig.~3c presents representative crystal structures generated by MatMind, spanning a broad compositional space from binary to quaternary systems and from fluorides to oxides. The generated structures include the rare-earth multinary fluoride SmTlCu$_2$F$_8$, the lithium rare-earth thiophosphate LiSmPS$_4$ (whose composition is closely related to lithium-ion solid-state electrolyte systems and carries potential functional materials significance~\cite{TODO_lithium_solid_state_electrolytes}), and the multinary alloy BaSn$_2$BiAu$_4$ featuring a long-period hexagonal superlattice structure. Particularly noteworthy is the case of Ti$_2$O$_3$: the training set contains four metastable polymorphs of Ti$_2$O$_3$ (space groups Pnma, R-3, P-31c, and P2$_1$/m, with $E_{\text{hull}}$ ranging from 0.007 to 0.079~eV/atom), yet the true thermodynamic ground state of Ti$_2$O$_3$---the corundum-type structure (space group R$\bar{3}$c, mp-458, $E_{\text{hull}}$~=~0.000~eV/atom)---never appears in the training data~\cite{TODO_thermodynamic_stability,TODO_ti2o3_corundum}. MatMind, having been exposed only to metastable polymorphs, generates from scratch a corundum-type configuration highly consistent with the experimentally known ground state~\cite{TODO_ti2o3_experimental_structure}. This result demonstrates that the model does not simply memorize structural templates from the training data, but rather establishes a deep understanding of crystal structure-stability relationships through the internalization of structure-activity knowledge, possessing the ability to infer thermodynamic ground-state configurations from metastable phase information.

\begin{figure}[htp]
    \centering
    \begin{adjustbox}{center}
        \includegraphics[width=18cm]{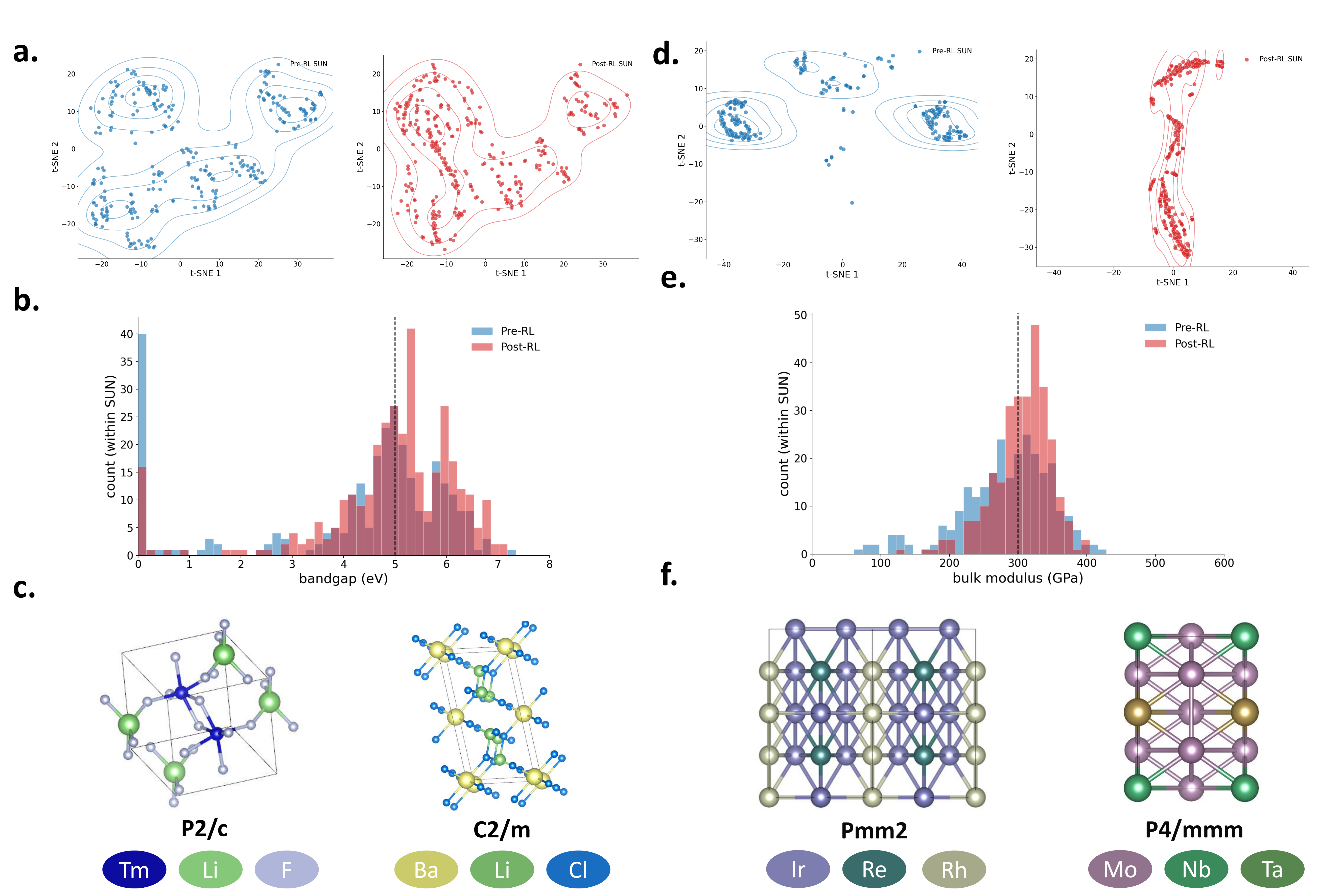}
    \end{adjustbox}
    \caption{Conditional crystal generation. \textbf{a,b}, Band gap conditional generation
(target $>5$~eV): t-SNE of generated S.U.N.\ structures before and after RL (\textbf{a}) and
the generated band gap distribution shifting toward the target (\textbf{b}); the satisfying
S.U.N.\ proportion rises from 18.0\% to 34.8\%. \textbf{d,e}, Bulk modulus conditional
generation (target $\sim 300$~GPa): t-SNE (\textbf{d}) and bulk modulus distribution
(\textbf{e}), with the satisfying S.U.N.\ proportion rising from 12.8\% to 26.8\%.
\textbf{c}, Representative band-gap-constrained structures (e.g., LiTmF$_4$, BaLi$_2$Cl$_4$).
\textbf{f}, Representative bulk-modulus-constrained structures (e.g., the platinum-group
alloys ReIr$_5$Rh$_2$ and ReIrPt$_6$).}
    \label{fig:Method}
\end{figure}

\subsubsection{Conditional crystal generation}
Building on the unconditional generation capability, MatMind further translates structure--activity reasoning capability into controllable inverse design capability (Fig.~4)~\cite{TODO_inverse_materials_design,TODO_conditional_crystal_generation}. The data sources for the conditional generation framework rely entirely on existing band gap and bulk modulus property annotations in the MP-20 database~\cite{TODO_conditional_crystal_generation,TODO_thermodynamic_stability}, requiring no additional DFT calculation investment~\cite{TODO_dft_materials_screening}. MatMind achieves property-guided inverse generation directly through its structure--activity understanding of existing data distributions. This characteristic endows the framework with inherent extensibility: while the current implementation supports conditional generation for band gap and bulk modulus, extending to additional properties requires only supplementary DFT annotations for the corresponding property, with no need to redesign the framework itself; any material property with accumulated large-scale DFT annotations can be directly integrated into MatMind's conditional generation pipeline without waiting for new computational data accumulation.
 
Conditional generation is an extremely challenging task, requiring the model to possess active inference capability for reverse-mapping property targets to structural space rather than relying solely on forward structure--activity understanding~\cite{TODO_inverse_materials_design,TODO_conditional_crystal_generation}. In the band gap conditional generation task (Fig.~4a, 4b), the training data contains approximately 6{,}000 samples satisfying the target band gap range ($>5$~eV), upon which the SFT stage has already established a reasonably strong generation capability; after RL, the band gap distribution of generated structures further concentrates toward the target value ($\sim 5$~eV), with the S.U.N. proportion rising from 18.0\% to 34.8\%. In the bulk modulus conditional generation task (Fig.~4d, 4e), training samples satisfying the target constraint ($\sim 300$~GPa) number only approximately 600, yet RL still drives a pronounced shift in the generated distribution: post-RL structures concentrate tightly around the target value, with the convergence trend in the t-SNE visualization being particularly striking~\cite{TODO_tsne}, and the S.U.N. proportion likewise achieving a substantial improvement (12.8\% to 26.8\%). This comparison demonstrates that physical reward signals can further strengthen the goal-directedness of generation on top of well-supported data foundations, while equally driving effective convergence of the generated distribution toward the target property interval under conditions of more limited data availability, reflecting the effective responsiveness of MatMind's conditional generation framework to physical reward signals~\cite{TODO_physics_informed_rl,TODO_rl_for_materials_generation}. Band gap and bulk modulus belong to physically distinct categories of electronic structure and mechanical properties, respectively~\cite{TODO_band_gap_electronic_structure,TODO_bulk_modulus_materials}; the consistent positive outcomes across both tasks indicate that the structure--activity representations established by MatMind generalize across property types rather than constituting local adaptation to a specific physical mechanism.
 
Fig.~4c presents representative generated structures satisfying the band gap constraint, including the rare-earth fluoride LiTmF$_4$ and the alkaline-earth--alkali-metal chloride BaLi$_2$Cl$_4$, both of which belong to wide-band-gap insulator families and exhibit local chemical environments highly consistent with known materials in the target band gap range~\cite{TODO_wide_band_gap_insulators,TODO_fluoride_chloride_insulators}. Fig.~4e presents representative generated structures satisfying the bulk modulus constraint, including the multinary platinum-group alloys ReIr$_5$Rh$_2$ and ReIrPt$_6$; the high electron density and strong metallic bonding characteristics of platinum-group elements endow such systems with outstanding mechanical rigidity, in strong physical self-consistency with the target high-bulk-modulus constraint~\cite{TODO_platinum_group_alloys,TODO_metallic_bonding_bulk_modulus}.


\subsection{Small-data-driven magnetization density property transfer}
\begin{figure}[htp]
    \centering
    \begin{adjustbox}{center}
        \includegraphics[width=18cm]{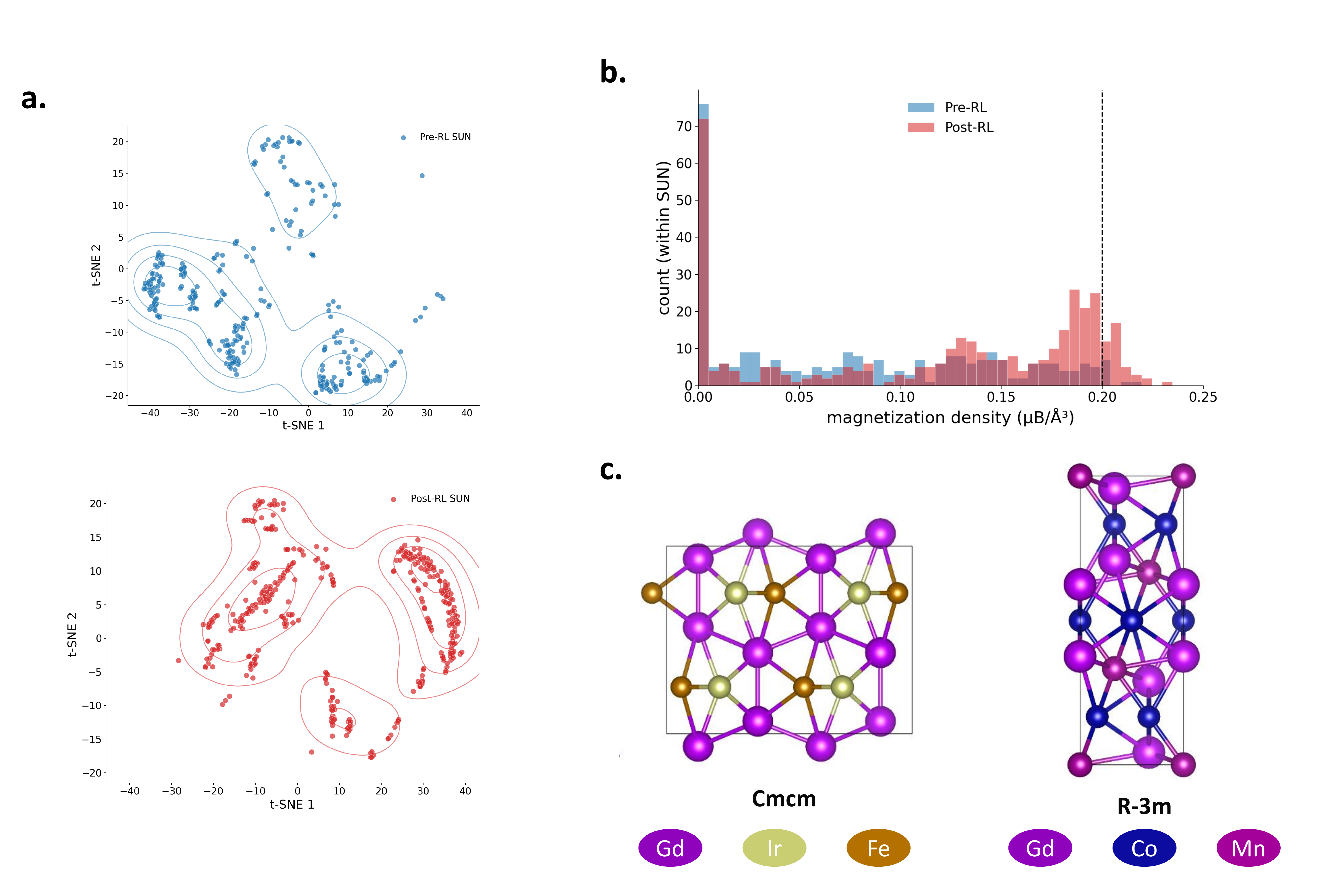}
    \end{adjustbox}
    \caption{Small-data magnetization density property transfer. With only 21 positive samples
($\geq 0.2~\mu_{\rm B}/\text{\AA}^3$) among over 600{,}000 training entries, physics-informed
RL increases the fraction of S.U.N.\ candidates satisfying the constraint from 6/500 (1.2\%)
to 26/500 (5.2\%). \textbf{a}, Magnetization density distribution of generated structures
before and after RL; the post-RL distribution develops a pronounced peak near the target
($0.2~\mu_{\rm B}/\text{\AA}^3$). \textbf{b}, t-SNE of S.U.N.\ structures in chemical space,
showing post-RL structures occupying regions distinct from the pre-RL clusters. \textbf{c},
Representative high-magnetization-density structures Gd$_2$FeIr (Cmcm) and Gd$_2$MnCo$_3$
(R$\bar{3}$m), combining rare-earth (Gd) and 3d transition-metal sublattices.}
    \label{fig:Method}
\end{figure}
High magnetization density materials are of central importance for permanent magnets, magnetic storage, and spintronic devices~\cite{TODO_permanent_magnets,TODO_magnetic_storage,TODO_spintronics}, yet materials combining high magnetization density with thermodynamic stability are sparsely represented in existing databases~\cite{TODO_thermodynamic_stability,TODO_data_scarcity_materials_discovery}. Within over 600{,}000 collected magnetic material entries, only 21 satisfy the target magnetization density constraint ($\geq 0.2~\mu_{\rm B}/\text{\AA}^3$), corresponding to a positive-sample fraction of approximately $3\times 10^{-5}$. This is approximately four orders of magnitude lower than the positive-sample fractions of the band gap and bulk modulus conditional generation tasks (18.0\% and 12.8\%, respectively). Property-guided generation under this regime therefore constitutes a stringent test of whether structure--activity reasoning has been internalized at a level that permits generalization beyond the immediate support of labeled positives~\cite{TODO_low_data_materials_discovery,TODO_inverse_materials_design}.

MatMind addresses this task through a two-stage training framework. The first stage applies property-guided supervised fine-tuning on the full corpus of magnetic material data (over 600{,}000 entries), establishing prior understanding of the structural distribution of magnetic materials and biasing the generative distribution toward chemical compositions and space groups associated with higher magnetization density. The second stage applies physics-informed reinforcement learning on top of the SFT model, using magnetization density as the reward signal for online optimization~\cite{TODO_physics_informed_rl,TODO_rl_for_materials_generation}. Both stages rely entirely on existing computational results from available databases, requiring no additional DFT calculation investment~\cite{TODO_dft_materials_screening,TODO_data_scarcity_materials_discovery}. Under the 21-positive-sample condition, supervised fine-tuning is fundamentally constrained by the distributional coverage of the available positive samples and cannot reliably generate structures within the target property interval~\cite{TODO_few_shot_materials_learning,TODO_data_imbalance_materials}. Physics-informed reinforcement learning circumvents this constraint by leveraging the computable physical reward to guide policy updates beyond the coverage of the 21 labeled positives, decoupling generative optimization from the scale of available positive supervision~\cite{TODO_physical_reward_guided_generation}.

Quantitatively, among 500 generated S.U.N. candidates, the number of structures satisfying the magnetization density constraint increases from 6 (1.2\%) before RL to 26 (5.2\%) after RL, corresponding to an approximately 4-fold improvement. For comparison, on the band gap task RL improves the satisfying-S.U.N. fraction from 18.0\% (90/500) to 34.8\% (174/500), and on the bulk modulus task from 12.8\% (64/500) to 26.8\% (134/500). The multiplicative gains across the three tasks are of comparable magnitude despite the positive-sample availability differing by approximately four orders of magnitude, indicating that physics-informed reinforcement learning provides effective property-guided optimization across data regimes spanning from dense to extremely sparse supervision. The magnetization density distribution of generated structures (Fig.~5a) shows that the Post-RL distribution develops a pronounced peak in the vicinity of the target value ($0.2~\mu_{\rm B}/\text{\AA}^3$), whereas the Pre-RL distribution remains broadly distributed in the low-magnetization regime, reflecting the distributional bias of the training corpus. The t-SNE visualization of S.U.N. structures in chemical space (Fig.~5b) shows that Post-RL structures occupy regions distinct from the Pre-RL clusters, indicating that RL drives the model into chemical regions not covered by the labeled positive samples rather than concentrating generation around known anchors~\cite{TODO_tsne,TODO_chemical_space_exploration}.

Fig.~5c presents representative high-magnetization-density structures generated by MatMind, including the rare-earth--3d--5d intermetallic Gd$_2$FeIr and the rare-earth--3d multinary intermetallic Gd$_2$MnCo$_3$. Both structures combine rare-earth elements (Gd) with 3d transition metals (Fe, Mn, Co), consistent with the design principle of established permanent-magnet families such as SmCo$_5$ and Nd$_2$Fe$_{14}$B~\cite{TODO_smco5_permanent_magnet,TODO_nd2fe14b_permanent_magnet}, in which the rare-earth sublattice contributes magnetocrystalline anisotropy while the 3d transition-metal sublattice provides high magnetic moment density~\cite{TODO_rare_earth_transition_metal_magnets,TODO_magnetocrystalline_anisotropy}. The emergence of such chemically meaningful compositions in a setting where only 21 positive samples were available indicates that physics-informed reinforcement learning enables the model to identify chemical environments associated with high magnetization density rather than interpolate among the labeled positive samples, providing structural-level plausibility for the magnetic properties of the candidate materials.

The magnetization density transfer experiment demonstrates the value of MatMind's physics-informed feedback framework from two perspectives. First, under the condition that only 21 positive samples exist within over 600{,}000 training entries, physics-informed reinforcement learning still yields a multiplicative improvement in target-property generation comparable in magnitude to that observed on tasks with four orders of magnitude more positive samples, indicating that the generative performance of the framework is not bottlenecked by the scale of available positive supervision. Second, this capability generalizes directly to other data-scarce functional properties: provided that a computable physical reward signal is available, MatMind's two-stage framework can transfer structure--activity knowledge from existing databases to directional generation tasks for new properties without incurring additional large-scale DFT calculation costs, providing a scalable and general path for computational discovery of functional materials whose target properties are represented by only a small number of known examples~\cite{TODO_low_data_materials_discovery,TODO_active_learning_materials,TODO_high_throughput_materials_discovery}.

\section{Methods}\label{sec3}


\subsection{Dataset preparation}

\subsubsection{Pretraining dataset}
The MatMind pretraining corpus is sourced from the thermodynamically stable subset of the Alexandria database~\cite{TODO_alexandria_database}. And stability threshold is $E_{\text{hull}} \leq$ 0.1~eV/atom~\cite{TODO_thermodynamic_stability}. The stable subset is selected rather than the complete database to ensure the reliability of property annotations: unstable structures exhibit high noise in computed properties due to geometric instability, making them unsuitable as a basis for structure-activity relationship learning~\cite{TODO_unstable_structures_property_noise}. Each crystal is incorporated in three complementary representations: CIF representation including chemical formula, space group, and Wyckoff positions, directly encoding crystal symmetry and atomic arrangement information~\cite{TODO_cif_format,TODO_space_group_symmetry,TODO_wyckoff_positions}; physical property annotations including DFT-computed band gap, bulk modulus, and formation energy, providing quantitative correspondence between structure and function~\cite{TODO_dft_materials_screening}; and natural language crystal descriptions automatically generated via RoboCrystallographer~\cite{TODO_robocrystallographer}, bridging structural information with the language representation space. The three data types are arranged in randomly interleaved sequences rather than organized by category. The core motivation for this design is that block-organized data would lead the model to independently learn each representation type without establishing cross-modal associations; random interleaving forces the model to simultaneously process structural, property, and descriptive information within a limited context window, naturally encoding bidirectional mappings among the three.

\subsubsection{Structure-activity relationship fine-tuning dataset}
The SAR enhanced fine-tuning dataset is sourced from the evaluation subset of the Alexandria database~\cite{TODO_alexandria_database}, ensuring distributional consistency with the pretraining data while strictly avoiding data leakage---the pretraining and fine-tuning subsets are completely non-overlapping at the structural level. The dataset comprises three task types in equal proportion (1:1:1) with 30000 total training samples (approximately 10000 per task type). Crystal performance ranking tasks provide multiple crystal structures and a target property, requiring the model to output the correct ordering by property value, assessing understanding of relative structure-activity relationships. Performance interval prediction tasks provide a single crystal structure and a property range, requiring the model to determine whether the property falls within the specified range, assessing quantitative perception of absolute structure-activity relationships. Target-guided crystal selection tasks provide a candidate crystal set and target property requirements, requiring the model to identify satisfying crystals, assessing the ability to apply structure-activity knowledge for inverse screening~\cite{TODO_inverse_materials_design}. Unlike data construction approaches relying on automatic generation or large language model distillation~\cite{TODO_llm_data_distillation}, each sample is manually annotated and quality-reviewed by materials science domain experts to ensure the scientific accuracy of reasoning chains. Each sample uses a structured instruction with explicit task requirements as input, with chain-of-thought reasoning as an intermediate bridge~\cite{TODO_chain_of_thought}: the CoT process explicitly describes the reasoning steps from crystal structure through analysis of key structural features to the final property judgment, enabling the model to learn not only correct outputs but also the scientific reasoning process of deriving answers along the structure-activity chain.

\subsubsection{Predictive task dataset}
Quantitative property prediction is evaluated on the Materials Project MP-20 dataset comprising 45,231 experimentally validated inorganic crystal structures with DFT-computed properties covering band gap, bulk modulus, formation energy, and other core material properties~\cite{TODO_conditional_crystal_generation,TODO_thermodynamic_stability}. MP-20 is selected over Alexandria as the evaluation benchmark both to maintain comparability with existing literature standards and because its predominance of experimentally validated structures provides a more reliable measure of model generalization to real materials prediction tasks~\cite{TODO_crystal_property_prediction_benchmark}. Language head training data is constructed from the MP-20 training set through a two-phase annotation and distillation process: domain experts first manually annotate a subset of samples to generate reference answers with structure-activity reasoning processes; these are then used as supervision signals to extend through distillation to the training samples with complete reasoning chains~\cite{TODO_reasoning_distillation}.

\subsubsection{Generative task dataset}
The denovo supervised fine-tuning dataset is sourced from a combined dataset of MP-20 and the Alexandria stable subset~\cite{TODO_conditional_crystal_generation,TODO_alexandria_database}, encompassing unconditional and conditional generation tasks. Unconditional generation tasks use pure crystal structure sequences as targets without any property constraints, enabling the model to establish broad prior distributions over the crystallographic chemical space~\cite{TODO_conditional_crystal_generation}. Conditional generation tasks encode target properties---including band gap and bulk modulus---as natural language instructions prepended to the input sequence, requiring the model to generate structures satisfying specified property constraints~\cite{TODO_conditional_crystal_generation}. MP-20 provides high-quality supervision signals from experimentally validated structures, while the Alexandria stable subset supplements with broader chemical space coverage from computationally validated structures; their combination enables the model to learn more comprehensive crystallographic chemical patterns while avoiding over-fitting to the distributional biases of experimental databases. All crystal structures are encoded using Wyckoff representation~\cite{TODO_wyckoff_positions}. Given a crystal structure $C$, its Wyckoff representation is defined as:
\begin{equation}
  W(C) = \bigl(g,\,\{(w_k, a_k, x_k)\}_{k=1}^{K}\bigr),
  \label{eq:wyckoff}
\end{equation}
where $g \in \{1, \ldots, 230\}$ is the space group number in the International Crystallographic Tables~\cite{TODO_international_crystallographic_tables}, $w_k$ is the Wyckoff letter of the $k$-th inequivalent site, $a_k$ is the atomic species at that site, $x_k \in [0,1)^3$ is the fractional coordinate, and $K$ is the total number of inequivalent sites. Compared with encoding using complete CIF sequences, Wyckoff representation collapses equivalent atomic sites into inequivalent sites through space group symmetry, substantially reducing sequence length while preserving complete crystallographic symmetry information~\cite{TODO_wyckoff_positions,cao2025space}.

\subsection{Model development methodology}

\subsubsection{Foundation model construction}
MatMind builds upon S1 based 8B as the base model~\cite{TODO_s1_base}, developing MatMind SAR through crystal scientific data alignment pretraining and structure-activity relationship enhanced fine-tuning. The pretraining stage uses standard autoregressive language modeling with cross-entropy loss to optimize next-token prediction probability over the three randomly interleaved crystal data types~\cite{TODO_autoregressive_language_modeling}; the SAR enhanced fine-tuning stage uses CoT-annotated data as supervision signals~\cite{TODO_chain_of_thought} to specifically strengthen explicit causal reasoning about structure-activity relationships while preserving the crystallographic understanding established during pretraining.

\subsubsection{Predictive model construction}
The predictive model is built upon MatMind SAR through a two-step training strategy, designed to resolve the inherent incompatibility between language modeling objectives and quantitative numerical prediction---language heads optimize token probability prediction and are inherently unsuited for precise continuous-value regression~\cite{TODO_tokenization_numerical_precision,TODO_joint_language_regression}. In the first step, all parameters $\theta_{\text{LLM}}$ of MatMind SAR are frozen and only the newly introduced numerical regression head undergoes warm-up training. The regression head performs mean pooling over all token hidden states of the final layer followed by a linear transformation, directly outputting numerical predictions in continuous space:
\begin{equation}
  \hat{y} = W_{\text{reg}} \cdot \frac{1}{L}\sum_{i=1}^{L} h_i + b_{\text{reg}},
  \label{eq:regression_head}
\end{equation}
where $h_i \in \mathbb{R}^d$ is the hidden state vector of the $i$-th token at the final layer, $L$ is the input sequence length, $W_{\text{reg}} \in \mathbb{R}^{1 \times d}$ is the linear projection matrix. The warm-up phase optimizes the regression head parameters using mean squared error loss:
\begin{equation}
  \mathcal{L}_{\text{pre}} = \frac{1}{N}\sum_{n=1}^{N}(\hat{y}_n - y_n)^2,
  \label{eq:loss_pre}
\end{equation}
where $N$ is the number of training samples and $y_n$ is the DFT reference value of the $n$-th sample. The purpose of the warm-up phase is to complete initial alignment between the regression head's parameter space and the frozen LLM representation space, preventing gradient noise from random initialization from disrupting the representation structure during joint training.

In the second step, all parameters of MatMind SAR are unfrozen and the language head and numerical regression head are jointly trained under a unified loss function~\cite{TODO_multitask_learning,TODO_joint_language_regression}:
\begin{equation}
  \mathcal{L}_{\text{all}} = \alpha \cdot \mathcal{L}_{\text{kno}} + \beta \cdot \mathcal{L}_{\text{pre}},
  \label{eq:loss_all}
\end{equation}
where $\mathcal{L}_{\text{kno}}$ is the autoregressive cross-entropy loss of the language head, supervised by the SAR reasoning distillation dataset~\cite{TODO_reasoning_distillation}:
\begin{equation}
  \mathcal{L}_{\text{kno}} = -\frac{1}{T}\sum_{t=1}^{T} \log p_\theta(x_t \mid x_{<t}).
  \label{eq:loss_kno}
\end{equation}
Here $T$ is the total number of tokens in the target sequence (including the complete CoT reasoning chain), and $p_\theta(x_t \mid x_{<t})$ is the model's predicted probability for the $t$-th token given the prefix. The weighting coefficients $\alpha$ and $\beta$ are determined by monitoring the Pareto frontier of language reasoning accuracy and property prediction MAE on the validation set~\cite{TODO_pareto_multiobjective_optimization}.

\subsubsection{Generative model construction}
The generative model is built through two steps. The first step applies supervised fine-tuning of MatMind SAR on the de novo instruction dataset, establishing basic capability for crystal structure sequence generation from scratch~\cite{TODO_supervised_fine_tuning,TODO_conditional_crystal_generation}. The supervised fine-tuning objective maximizes the conditional log-likelihood of crystal structure Wyckoff sequences given conditional input $\mathbf{c}$~\cite{TODO_wyckoff_positions}:
\begin{equation}
  \mathcal{L}_{\text{SFT}} = -\mathbb{E}_{(C, \mathbf{c}) \sim \mathcal{D}_{\text{SFT}}} \left[ \sum_{t=1}^{T} \log p_\theta(x_t \mid x_{<t}, \mathbf{c}) \right],
  \label{eq:loss_sft}
\end{equation}
where $\mathbf{c} = \emptyset$ for unconditional generation and $\mathbf{c}$ contains natural language descriptions of target properties for conditional generation; $x_t$ is the $t$-th token of the $W(C)$ sequence; $T$ is the total sequence length.

The second step applies physics-informed reinforcement learning via Group Relative Policy Optimization (GRPO)~\cite{TODO_physics_informed_rl,TODO_grpo}. For the $G$ candidate structures $\{C_i\}_{i=1}^{G}$ generated by the policy model $\pi_\theta$ at each step, GRPO estimates the advantage of each structure through within-group relative comparison. The within-group relative advantage is defined as:
\begin{equation}
  A_i = \frac{R(C_i) - \mu_R}{\sigma_R + \epsilon},
  \label{eq:advantage}
\end{equation}
where $\mu_R = \frac{1}{G}\sum_{i=1}^{G} R(C_i)$ is the within-group reward mean, $\sigma_R$ is the within-group reward standard deviation, and $\epsilon$ is a numerical stability constant. The GRPO policy update objective is:
\begin{equation}
  \mathcal{L}_{\text{GRPO}}(\theta) = -\mathbb{E}\left[\sum_{i=1}^{G} \frac{1}{G}
    \min\!\left(\frac{\pi_\theta(C_i)}{\pi_{\text{old}}(C_i)} A_i,\;
    \mathrm{clip}\!\left(\frac{\pi_\theta(C_i)}{\pi_{\text{old}}(C_i)},\,
    1{-}\epsilon,\, 1{+}\epsilon\right) A_i\right)\right]
    + \beta \cdot D_{\mathrm{KL}}[\pi_\theta \| \pi_{\text{ref}}],
  \label{eq:grpo}
\end{equation}
where $\frac{\pi_\theta(C_i)}{\pi_{\text{old}}(C_i)}$ is the importance sampling ratio; the clip operation restricts the importance ratio to $[1-\epsilon, 1+\epsilon]$ to prevent training instability~\cite{TODO_ppo}; and the KL divergence regularization term with coefficient $\beta$ constrains the deviation between the current policy $\pi_\theta$ and the reference policy $\pi_{\text{ref}}$ (the initial model after supervised fine-tuning)~\cite{TODO_supervised_fine_tuning}.

\subsection{Reward function design}

\subsubsection{Validity reward}
The validity gate $V(C)$ is a prerequisite for total reward computation, implemented
through the logical AND of three physical constraints:
\begin{equation}
  V(C) = V_{\text{dist}}(C) \cdot V_{\text{charge}}(C) \cdot V_{\text{relax}}(C).
  \label{eq:validity}
\end{equation}
The interatomic distance constraint $V_{\text{dist}}(C) = \mathbf{1}[\min_{i \neq j} d_{ij} \geq d_{\min}]$
requires that the distance $d_{ij}$ between any two atoms satisfies $d_{\min} = 0.5$~\AA,
excluding non-physical structures with atomic overlap~\cite{TODO_multiobjective_materials_generation}.
The charge neutrality constraint $V_{\text{charge}}(C) = \mathbf{1}[\sum_i q_i = 0]$ requires
that the sum of all atomic oxidation states $q_i$ equals zero, ensuring generated structures
satisfy basic electrochemical conservation~\cite{TODO_charge_neutrality_crystals}. The
relaxation convergence constraint $V_{\text{relax}}(C) = \mathbf{1}[\text{MLIP relaxation converges without error}]$
requires that the NequIP-OAM-XL relaxation process completes successfully without triggering
numerical errors or abnormal termination~\cite{TODO_nequip,TODO_omat_reference,tan2026high，TODO_structure_relaxation}.
Failure of any single constraint results in $V(C) = 0$, zeroing all subsequent rewards for that
structure.

\subsubsection{Crystal embedding}
To support efficient computation of novelty and diversity rewards, a dual-descriptor embedding
is computed for each structure $C$ passing validity verification. The structural descriptor
$f_S(C) \in \mathbb{R}^{d_S}$, computed via CrystalNNFingerprint~\cite{TODO_crystalnn_fingerprint},
captures bonding patterns, coordination numbers, and bond length distributions in the local
coordination environment around each atom. The compositional descriptor
$f_E(C) \in \mathbb{R}^{d_E}$, computed via ElementProperty~\cite{TODO_elementproperty,TODO_crystalnn_fingerprint},
encodes elemental composition and stoichiometric ratio information. Together they constitute
the complete crystal embedding:
\begin{equation}
  f(C) = [f_S(C);\; f_E(C)] \in \mathbb{R}^{d_S + d_E}.
  \label{eq:embedding}
\end{equation}

\subsubsection{Stability reward}
The stability reward quantifies the thermodynamic stability of generated structures using the
energy above hull $E_{\text{hull}}(C')$ of the MLIP-relaxed structure $C'$, evaluated against
the Materials Project (MP-20) convex hull. The energy of each relaxed structure is computed
with the NequIP-OAM-XL potential, an OMat-family foundation potential, and is
calibrated using the OMat energy-correction scheme before $E_{\text{hull}}$ is evaluated. This
calibration places the MLIP total energies on a reference surface consistent with the MP hull,
yielding $E_{\text{hull}}$ estimates that are directly comparable across methods. A convex
decay mapping is then applied over the metastable window:
\begin{equation}
  R_S(C) = \begin{cases}
    1 & E_{\text{hull}}(C') \leq \tau_{\text{lo}} \\[6pt]
    \left(\dfrac{\tau_{\text{hi}} - E_{\text{hull}}(C')}{\tau_{\text{hi}} - \tau_{\text{lo}}}\right)^{\gamma}
      & \tau_{\text{lo}} < E_{\text{hull}}(C') < \tau_{\text{hi}} \\[10pt]
    0 & E_{\text{hull}}(C') \geq \tau_{\text{hi}}
  \end{cases}
  \label{eq:stability_reward}
\end{equation}
where $\tau_{\text{lo}} = 0.1$~eV/atom, $\tau_{\text{hi}} = 0.2$~eV/atom, and $\gamma = 2$.
The convex exponent ($\gamma > 1$) makes the reward rise steeply only as $E_{\text{hull}}$
approaches the stable boundary $\tau_{\text{lo}}$, concentrating optimization pressure on
near-stable structures rather than rewarding marginal gains in the high-energy regime.

\subsubsection{Novelty reward}
The novelty reward quantifies the degree to which a generated structure deviates from known
chemical space. The reference set $\mathcal{D}_{\text{ref}}$ is the combined MP-20 and
Alexandria stable-subset training corpus, treated as the representation of known
materials~\cite{TODO_chemical_space_exploration}. Novelty is assigned in two stages: a hard
structural-match gate, followed by a continuous fingerprint-distance score.

\emph{Hard novelty gate.} Each generated structure $C_i$ is first tested against
$\mathcal{D}_{\text{ref}}$ through a bucketed StructureMatcher comparison: reference structures
sharing the same coarse signature as $C_i$ are retrieved, and if any of them matches $C_i$
under StructureMatcher, $C_i$ is judged non-novel and its novelty reward is set to zero:
\begin{equation}
  m_i^{\text{nov}} = \mathbf{1}\!\left[\,\nexists\,C_j^* \in \mathcal{D}_{\text{ref}}
  : \mathrm{StructureMatcher}(C_i, C_j^*)\,\right].
  \label{eq:novelty_mask}
\end{equation}

\emph{Soft novelty score.} For structures passing the hard gate, the nearest-neighbor distance
to $\mathcal{D}_{\text{ref}}$ is computed independently in the structural and compositional
fingerprint spaces:
\begin{equation}
  d_S(C_i) = \min_{C_j^* \in \mathcal{D}_{\text{ref}}} \|f_S(C_i) - f_S(C_j^*)\|_2, \qquad
  d_C(C_i) = \min_{C_j^* \in \mathcal{D}_{\text{ref}}} \|f_E(C_i) - f_E(C_j^*)\|_2.
  \label{eq:novelty_dist}
\end{equation}
If both distances exceed their cutoffs ($d_S > \delta_S \wedge d_C > \delta_C$), the structure
receives the maximal soft score of $1$; otherwise each axis is mapped to $[0,1]$ through a
monotonic scoring function $s(\cdot;\delta)$ of the distance relative to its cutoff, and the two
axes are averaged:
\begin{equation}
  \tilde{R}_N(C_i) =
  \tfrac{1}{2}\bigl(s(d_S(C_i);\delta_S) + s(d_C(C_i);\delta_C)\bigr) \in [0,1].
  \label{eq:novelty_score}
\end{equation}
The final novelty reward combines the two stages:
\begin{equation}
  R_N(C_i) = \tilde{R}_N(C_i) \cdot m_i^{\text{nov}}.
  \label{eq:novelty_reward}
\end{equation}

\subsubsection{Diversity reward}
The diversity reward is based on the maximum entropy principle over the within-group structure
distribution~\cite{TODO_maximum_entropy_principle}. Pairwise distance matrices between all
structures within the group are first computed independently in structural and compositional
descriptor spaces:
\begin{equation}
  D_{ij}^S = \|f_S(C_i) - f_S(C_j)\|_2, \qquad
  D_{ij}^C = \|f_E(C_i) - f_E(C_j)\|_2.
  \label{eq:diversity_dist}
\end{equation}
Distances are converted to similarity logits using temperature parameter $\tau$:
\begin{equation}
  \tilde{Z}_{ij} = -\frac{D_{ij}}{\tau}, \qquad \tilde{Z}_{ii} = -\infty.
  \label{eq:logits}
\end{equation}
Softmax probability distributions over the remaining structures are computed for each
structure $i$:
\begin{equation}
  P_{ij} = \frac{\exp(\tilde{Z}_{ij} - \max_k \tilde{Z}_{ik})}{\sum_{k \neq i} \exp(\tilde{Z}_{ik} - \max_k \tilde{Z}_{ik})}.
  \label{eq:softmax}
\end{equation}
Shannon entropy is computed for each structure and normalized by the maximum entropy
$\log(N-1)$~\cite{TODO_shannon_entropy}:
\begin{equation}
  H_i = -\sum_{j \neq i} P_{ij} \log P_{ij}, \qquad
  \hat{H}_i = \frac{H_i}{\log(N-1)} \in [0, 1].
  \label{eq:entropy}
\end{equation}
This process is performed independently in structural and compositional spaces, yielding
$\hat{H}_i^S$ and $\hat{H}_i^C$. A within-group uniqueness gate identifies duplicate structures
using a StructureMatcher comparison augmented with Hume-Rothery substitution rules, under which
two structures are treated as equivalent if one maps onto the other through element
substitutions between species of comparable radius, electronegativity, and oxidation state:
\begin{equation}
  \mathrm{dup}_i = \mathbf{1}\!\left[\exists\, j \neq i : \mathrm{StructureMatcher}_{\text{HR}}(C_i, C_j)\right], \qquad
  m_i^{\text{uniq}} = 1 - \mathrm{dup}_i.
  \label{eq:uniq_mask}
\end{equation}
The final diversity reward is:
\begin{equation}
  R_U^{(i)} = \frac{\hat{H}_i^S + \hat{H}_i^C}{2} \cdot m_i^{\text{uniq}}.
  \label{eq:diversity_reward}
\end{equation}

\subsubsection{Compositional ratio re-balancing}
To discourage the policy from collapsing onto a small set of stoichiometric templates, we apply
a within-group ratio re-balancing factor that operates at the level of composition
stoichiometry, complementing the descriptor-based diversity reward. For each structure we form
a stoichiometric-ratio fingerprint by normalizing the reduced-composition ratios, rounding to
one decimal place, and sorting. The empirical frequency of each fingerprint within the rollout
group is estimated; a structure whose fingerprint frequency exceeds a high threshold is
down-weighted by $\rho_{\text{pen}} = 0.85$, while a structure carrying a rare fingerprint is
up-weighted by $\rho_{\text{bonus}} = 1.15$, yielding a multiplicative factor $m_i^{\text{ratio}}$.

\subsubsection{Stochastic element conditioning}
To support controllable composition, with probability $p_{\text{elem}}$ each rollout is assigned
a target element drawn from a fixed pool. A generated structure satisfies the constraint
($m_i^{\text{elem}} = 1$) only if it contains the assigned target element, and is otherwise
assigned zero reward; unconstrained rollouts have $m_i^{\text{elem}} = 1$ by default. In the
unconditional generation experiments reported here $p_{\text{elem}} = 0$, so this module is
inactive; it is enabled only for composition-controlled generation.

\subsubsection{Total reward}
After passing the validity gate, the total reward combines stability, novelty, and diversity
(and, for property-conditioned runs, a property-target term $R_P$), modulated by the
ratio-rebalancing and element-conditioning factors:
\begin{equation}
  R(C_i) = \mathbf{1}[V(C_i) = 1] \cdot m_i^{\text{ratio}} \cdot m_i^{\text{elem}}
  \cdot \bigl(R_S(C_i) + R_N(C_i) + R_U^{(i)} + R_P(C_i)\bigr).
  \label{eq:total_reward}
\end{equation}
The hard novelty gate $m_i^{\text{nov}}$ and the within-group uniqueness gate $m_i^{\text{uniq}}$
are folded into $R_N$ and $R_U^{(i)}$ respectively (Eqs.~\eqref{eq:novelty_reward} and
\eqref{eq:diversity_reward}). For unconditional generation $R_P = 0$. All reward components are
mapped to $[0,1]$; the additive form ensures that a zero in any single objective still leaves
informative gradients from the others, while the multiplicative factors act as filters that
remove duplicated or constraint-violating structures from optimization. Within-group
standardization is applied to $R(C_i)$ before the GRPO advantage computation of
Eq.~\eqref{eq:advantage}.

\subsection{Evaluation metrics}

\subsubsection{Property prediction}
Quantitative property prediction is evaluated using mean absolute error (MAE)~\cite{TODO_mean_absolute_error}. MAE directly reflects the average deviation between predicted and reference values:
\begin{equation}
  \text{MAE} = \frac{1}{N}\sum_{n=1}^{N}|\hat{y}_n - y_n|.
  \label{eq:mae}
\end{equation}

convergence~\cite{TODO_multiobjective_materials_generation}.

\subsubsection{Crystal generation}
Crystal generation is evaluated using the Stable-Unique-Novel (S.U.N.) rate as a composite
metric~\cite{TODO_conditional_crystal_generation}:
\begin{equation}
  \text{S.U.N} = \frac{|\{C : E_{\text{hull}}(C') \leq 0 \wedge C \notin \mathcal{D}_{\text{ref}} \wedge C \in \mathcal{U}\}|}{N_{\text{total}}}.
  \label{eq:sun}
\end{equation}
For stability evaluation, generated structures are relaxed with the eSEN-30M-OAM\cite{fu2025learningsmoothexpressiveinteratomic}
potential, an OMat-family foundation potential, and their energies are
calibrated using the OMat energy-correction utility before $E_{\text{hull}}$ is computed against
the MP-20 convex hull~\cite{TODO_omat_reference,TODO_energy_correction_mlip}. The OMat energy
correction places all relaxed energies on an MP-consistent reference surface, which we adopt as
the protocol that yields the most reliable $E_{\text{hull}}$ estimates in our setting. All stages
of MatMind rely on OMat-family potentials with OMat energy correction, placing the training
reward and the evaluation metric within a consistent energy-reference framework. We deliberately
use a different OMat-family potential for evaluation (eSEN-30M-OAM) from the one used in the
reinforcement-learning reward and the validity-gate relaxation (NequIP-OAM-XL): because the
evaluation potential is not the one optimized against during training, the reported S.U.N.\ rate
reflects stability that generalizes across independent potentials rather than overfitting to the
reward model. To ensure a fair comparison, the baseline methods (MatterGen, DiffCSP) are
evaluated under this identical relaxation, energy-correction, and MP-hull protocol. The stability
condition requires $E_{\text{hull}}(C') \leq 0$~eV/atom after relaxation; the novelty condition
$C \notin \mathcal{D}_{\text{ref}}$ requires that $C$ does not match any structure in the combined
MP-20 + Alexandria stable-subset training corpus under the StructureMatcher criterion of
Eq.~\eqref{eq:novelty_mask}; the uniqueness condition $C \in \mathcal{U}$ requires structural
distinctness within the generated set; and $N_{\text{total}}$ is the total number of generated
structures. Validity is defined as the proportion of generated structures simultaneously
satisfying the three constraints of interatomic distance, charge neutrality, and relaxation
convergence~\cite{TODO_multiobjective_materials_generation}.

\section{Conclusion}\label{sec4}

We have introduced MatMind, a generative foundation model for crystal materials science developed under the LLM-based paradigm through the coordinated activation of structure--activity knowledge and physics-informed feedback. MatMind achieves the lowest mean absolute error against both graph neural network and language-model-based specialists on the prediction of energy above hull, bulk modulus, and band gap, reaches an S.U.N.\ rate of 65.3\% on unconditional crystal generation relative to mainstream diffusion and language-model baselines, and drives generated distributions toward target intervals across conditional generation tasks, including magnetization-density-conditioned generation, where a comparable multiplicative improvement is obtained with only 21 positive samples available within over 600{,}000 training entries. These results show that, when supplied with sufficient materials-specific prior and grounded by physical feedback, an LLM-based foundation can be specialized into competitive predictive and generative capabilities for crystal materials science, complementing the narrow architectures that dominate the field today. MatMind thereby provides a foundation on which property prediction, structure generation, and structure--activity reasoning can be jointly developed, and on which subsequent extensions to broader property coverage, multi-property conditional design, and closer integration with experimental and synthesis workflows can be systematically pursued.



\bibliography{sn-bibliography}

\end{document}